# Promoting the linguistic diversity of TEI in the Maghreb and the Arab region

Henri Hudrisier[1], Rachid Zghibi[2], Sihem Zghidi[3], Mokhtar Ben Henda[4]

1 : Université Paris 8, Labo Paragraphe, Chaire Unesco-Iten, France
2 : Institut Supérieur de Documentation, Université La Manouba, Tunisie
3 : Institut Supérieur de Documentation, Université La Manouba, Tunisie
4 : Université Bordeaux Montaigne, Labo MICA UR 4426, Chaire Unesco-Iten, France

## Presentation

Since many centuries, the Maghreb region is experiencing significant linguistic hybridization that slowly impacts on its cultural heritage. Besides Libyan, Latin and Ottoman contributions, significant other amounts of resources in various cultures and languages have been accumulated in the Maghreb region, either derived from classical Arabic (i.e. regional dialects) or from various dialects of Berber (i.e. Kabyle). Several resources are even composed simultaneously in several common or restricted languages (literary Arabic, colloquial Arabic, French, English, Berber) like newspapers, "city printing", advertising media, popular literature, tales, manuals for learning languages, etc. These resources are often written in a hybrid script mixing both classical and vernacular Arabic, or combining transliteration forms between Latin, Arabic and Tifinagh (traditional Berber script). Unlike many traditional textual resources (conventional printed documents and medieval manuscripts), it does not exist today vast corpora of texts in vernacular idioms and scripts. But our hypothesis is that the growing awareness of the diversity of these textual resources would rapidly result in an exponential increase of the number of researchers interested in collecting and studying classical old texts and oral resources. The standard TEI encoding format provides in this respect a unique opportunity to optimize these resources by ensuring their integration into the international cultural heritage and their use with maximum technical flexibility. The "HumanitéDigitMaghreb" project, which is the subject of this intervention, intents to address several aspects of theses research objectives and to initiate their appropriation.

## Research hypothesis

The project targets both oral corpus and the rich text resources written in the Maghreb region. It focuses particularly on the continuity, for more than 12 centuries, of a classical still alive Arabic language and on the extreme hybridization of vernacular languages sustained by the rich Libyan, Roman, Hebrew and Ottoman influences and by the more recent French, Spanish and Italian linguistic interference. In short, the Maghreb is a place of extremely abundant, but much unexploited, textual studies.

Our project permits comparative visions to understand how to transform TEI - originally designed for classical and modern European languages (Latin, medieval languages, etc. ...) - in order to work on corpora in literary Arabic and in mixed languages and scripts. For example, how researchers from the Maghreb, who invest in the French metric study and fully understand the TEI markup, can understand the subtlety of Arabic meter markup? How do they develop and give examples, when possible, of markup terminological equivalents of metric description in English, French and Arabic?
How can they see if there are really specific «Arabic» structural concepts and then provide the appropriate tags for them. These questions can concern "manuscripts", "critical apparatus", "performance text", etc...? For "TEI speech", we assume, however, that it is not really likely to be the specific method to apply although much work remains to be done. Doing this, we are aware that researches on similar adaptations are undertaken in other languages and cultures: Korean, Chinese, Japanese, etc. Theses adaptations and appropriations of the TEI experiences are of high interest for us.

## Core questions

As a starting point, we consider that the use of TEI in the Maghreb and the Middle East is still sporadic and unrelated. The existing work is mainly concentrated on the study of manuscripts and rare books. This focus can be explained primarily by the existence of large collections of Oriental manuscripts in western digital collections that are TEI encoded since a long time. It can also be explained by the urgency felt within the Arab cultural institutions to accelerate the preservation of cultural heritage from deterioration. Thus, we assume that TEI relatively profited from all experiences and projects for encoding Arabic manuscripts. However, this effort seemingly still needs a larger amount of feedbacks of other nature, generated from other types of resources with other forms of complexity (mainly linguistic and structural). The question that drives us here is to know how the complexity of that cultural heritage (that of the Maghreb as much as we are concerned) would be of any contribution to TEI? How to define its cultural and technological distinctiveness compared to the actual TEI-P5 and what are the solutions?

## Methodology

In the project "HumanitéDigitMaghreb", we particularly focus on the methods of implementing the TEI to address specific complex structures of multilingual corpus. We achieved some results, but on the long term, we especially concentrate on practical and prospective issues of very large standardized and linguistically structured corpora that will allow, for all linguistic communities (and we concentrate here on the Maghreb world), to constitute appropriate references in order to interact correctly with translation technologies and e-semantics in the future. On this last point, it is essential that the community of Arab and Berber researchers mobilize without delay to provide these languages (both written and oral) with their digital modernity. Three steps are to be taken in this respect:

1. The first step, which is beyond the limits of our project "HumanitéDigitMaghreb", inevitably involves a linguistic and sociocultural analysis of the Arabic context in order to clarify three points: first, how the TEI, in its current and future versions, would encode the Arab cultural heritage; second, how the Arabic context surpasses the limits of one level of standard cataloguing (MARC, ISBD, AACR2, Dublin Core) ; and third, how it succeeds to standardize the different approaches of its heritage scholarly reading.

2. In its constant evolution, and the need to strengthen its internationalization, the TEI community would undoubtedly profit from these cultural and linguistic characteristics. This would require also that this community be well organized to provide adequate encoding standardized formats for a wide range of linguistically-heterogeneous textual data. We can imagine here the encoding needs of electronic texts in Arabic dialects profoundly scattered with transliterated incises or written in different characters. These texts are potentially very complex. Besides connecting these materials to each other, like in parallel data (often bilingual), there are further levels of complexity inherent to the use of character sets and multiple non-standard transcription systems (different from the International Phonetic Alphabet) and related to the need of transcribing the speech in an overwhelmingly oral society, which poses interesting encoding problems. The second step, which is under the scope of our proposal, is to produce TEI standard references in local languages and to introduce them to academic and professional communities. These standards help address issues of specific linguistic complexity like hybridization of digital resources (local dialects) and preservation of a millenary oral and artistic heritage. Thus, the issue of character sets is not without consequence to represent local dialects, in large part because many of their cultural aspects were not taken into account in the development of existing standards (transcribing numbers and symbols, some forms of ligatures, diplomatic and former alphabets). There are, for example, many properties of the Arabic or Berber languages, as the tonal properties, regional synonymy and classical vocalization, (notarial writing) that require special treatment. Current standards, in particular the Unicode and furthermore ISO 8859 standards, do not take into account many of these aspects.

3. The third step, in which we are also engaged, is the creation of a community of practice specialized in the treatment of specific resources. We note here that most of these resources are potentially complex and certain features require probably specific markup arrangements. This means that a dynamic environment is required to specify the encoding of these documents - an environment in which it is easy to encode simple structures, but where more complex structures can be also encoded. Therefore, it is important to have specifications that can be easily extended when new and interesting features are identified.

We are interested in TEI not only for its collegial dynamics open on non-European linguistic diversity (Japan, China, Korea…), but also for its eclectic research disciplines (literature, manuscripts, oral corpus, research in arts, linguistics...) and its rigor to maintain, enrich and document open guidelines on diversity ensuring at the same time the interoperability of all produced resources.

## Results

The results of our work are reflected through a website that lists a collection of TEI encoded samples of resources in areas such as music, Arabic poetry, Kabyle storytelling and oral corpus. To achieve this, we went through a fairly rapid first phase of TEI guidelines appropriation. The second phase would be a larger spreading of the TEI guidelines among a wider community of users including graduate students and mostly scholars not yet convinced of the TEI added-value in the Maghreb region. Those could be specialists of Arabic poetry, specialists of the Berber language, musicologists, storytelling specialists... The translation of the TEI P5 in French and Arabic, but also the development of a sample corpus and the construction of TEI multilingual terminology or glossary in English/French/Arabic, seems very necessary.

We also intend to propose research activities within other communities acting at national and regional levels in order to be in total synergy with the international dynamics of TEI. We have been yet involved in an international project, the "Bibliothèque Numérique Franco-Berbère" aimed at producing Franco-Berber digital resources with a funding from the French speaking international organization. In short, by getting engaged in the school of thought of Digital Humanities and TEI, we explicitly intend to give not only a tangible and digital reality to our work, but we try to make it easily cumulative, upgradable and exchangeable worldwide. More specifically, we expect that our work be easily exchangeable between us and our three Maghreb partner languages (Arabic, French, Berber) beside English.

Apart from the emerging issue of management and setting a standardized and interoperable digital heritage, it is obvious that specialists in this literary heritage should largely explore the methods of study and cataloguing. Therefore, this article is limited to discuss only questions of scholars and professionals (libraries and research centres) appropriation of digital humanities tools and services in the Oriental context. We will focus, among other issues, on compared cultural problems by facing European ancient manuscripts study to the Arabic cultural context.

# Promouvoir la diversité linguistique des TEI au Maghreb et dans la région arabe

## Constat

Depuis des siècles, la région du Maghreb connait une hybridation linguistique importante qui impacte lentement sur son patrimoine culturel. Au-delà des strates libyques, latines, ottomanes se sont accumulées des quantités non négligeables de ressources dans plusieurs versions linguistiques dérivées de l'arabe classique (dialectes régionaux) en plus des versions dialectales berbères et kabyles. Plusieurs ressources sont même composées simultanément en plusieurs langues d'usage courant ou réservé (arabe littéraire, arabe dialectal, français, anglais, berbère) comme les journaux, l'imprimerie de ville, les supports publicitaires, la littérature populaire, les contes, les manuels d'apprentissage des langues parlées, etc. Elles sont souvent écrites dans une graphie hybride mélangeant à la fois l'arabe classique et vernaculaire, ou des formes de translittération latin-arabe ou arabe-latin voire aussi en tifinagh (écriture berbère traditionnelle) notamment au Maroc. Contrairement à beaucoup de ressources textuelles classiques (documents conventionnels imprimés et manuscrits médiévaux), il n'y a pas de vastes corpus préexistants de textes en langues vernaculaires. Mais cette typologie de documents est susceptible d'augmenter de façon exponentielle dans un proche avenir et nombre de chercheurs s'intéressent au recueil des corpus oraux et à l'étude de ces littératures orales plurielles. Cela présente une opportunité unique pour créer ces ressources en utilisant un format de codage standard comme la TEI qui peut assurer leur intégration dans le patrimoine culturel général et leur réutilisabilité avec une flexibilité technique maximale. Le projet « HumanitéDigitMaghreb » qui fait l'objet de cette intervention, s'intéresse à plusieurs aspects de ces questions de recherche.

## Questions de recherche

Le projet cible à la fois les corpus oraux mais aussi les riches ressources écrites dans la région du Maghreb. Il se focalise notamment sur la permanence sur 12 siècles d'un arabe classique aujourd'hui encore vivant et sur l'extrême hybridation des langues vernaculaires entretenues par la richesse des sources libyques, romaines, hébraïques, ottomanes auxquelles s'ajoutent encore les ressources françaises, espagnoles et italiennes. Bref, le Maghreb est un lieu d'études textuelles extrêmement foisonnant et très inexploité.

Notre projet nous permet des visions comparées qui permettent de comprendre comment transformer la TEI adaptée aux langues européennes et classiques (latin), pour travailler sur les corpus en arabe littéraire et en langues et écritures hybrides. Par exemple comment des chercheurs du Maghreb qui s'investissent dans l'étude métrique française et en comprennent complètement le balisage TEI, peuvent ensuite comprendre la subtilité du balisage métrique arabe. Comment développent-ils des exemples de balisage et donner, quand c'est possible, des équivalents terminologiques de la description métrique en anglais, français et arabe. Comment peuvent-ils voir s'il y a réellement des concepts structurels « spécifiquement arabe » et proposer alors les balises appropriées. Cet exemple peut se décliner autant pour les manuscrits que pour les apparats critiques, les « Performance text », etc. Pour la « TEI speech » nous supposons qu'il ne risque vraiment pas d'y avoir des spécificités de méthode même si beaucoup de travail reste à faire.

## Problématique

Or, l'usage de la TEI dans le Maghreb et le Moyen-Orient arabe est encore timide et disparate. Il est essentiellement concentré sur l'étude des manuscrits et des livres rares. Cette focalisation s'expliquerait d'abord par l'existence de grandes collections de manuscrits orientaux dans les fonds numériques occidentaux qui pratiquent depuis longtemps l'encodage TEI. Elle s'explique ensuite par l'urgence ressentie au sein d'instances culturelles arabes pour accélérer la préservation d'un patrimoine culturel en risque de délabrement. La TEI P5 a, de ce fait, relativement profité d'éclairages apportés par des expériences et projets arabes d'encodage de manuscrits, mais il lui manquerait un pan entier de feed-back d'autres natures inhérents à d'autres types de ressources et à d'autres formes de complexités (notamment linguistiques et structurelles) du patrimoine culturel maghrébin. La question qui nous mobilise ici est de savoir en quoi la complexité de ce patrimoine culturel serait-elle d'un quelconque apport à la TEI P5 ? Où se situent leur particularisme culturel et technique par rapport à la TEI et quelles en sont les solutions ?

## Méthodologie

Dans cet article, nous nous focalisons particulièrement sur les méthodes de mise en œuvre de la TEI pour traiter des cas spécifiques de la structuration de corpus multilingues. Nous donnerons les résultats de nos travaux, mais nous examinerons aussi les enjeux concrets et prospectifs de très grands corpus linguistiquement structurés de façon normative qui permettront de constituer pour toute communauté linguistique (et pour ce qui nous concerne le monde maghrébin) le référentiel de contexte indispensable pour interagir avec les outils de traductique et d'e-sémantique du futur. Sur ce dernier point, tant pour les corpus écrits que pour les corpus oraux (versus sonore ou versus transcription), il est indispensable que la communauté des chercheurs arabes et berbères se mobilisent pour parvenir sans retard à doter ces langues de leur modernité digitale. Pour y parvenir, trois stades sont à franchir.

1. Un premier stade - au-delà des limites de notre papier - passe inéluctablement par un travail d'analyse linguistique et socioculturel du contexte arabophone pour définir comment la TEI, dans ses versions actuelles et futures, permettrait d'encoder le patrimoine culturel arabe, de dépasser le seul niveau de son catalogage par les normes en vigueur (MARC, ISBD, AACR2, Dublin Core) et surtout de normaliser les différentes approches de sa lecture savante. Dans son évolution constante, et le besoin de renforcer sa facette d'internationalisation, la communauté TEI s'enrichirait sans doute de ces spécificités culturelles et linguistiques, ce qui exigerait bien sûr que cette même communauté s'organise pour proposer des formats de codage standard adéquats pour représenter un large éventail de données textuelles linguistiquement hétérogènes. Nous pouvons imaginer ici les besoins d'encodage de textes électroniques en dialectes arabes profondément parsemés d'incises translittérées ou en caractères différents. Ces textes sont potentiellement très complexes. En plus de relier très souvent ce matériel à d'autres, éventuellement des données parallèles (bilinguisme fréquent), ils constituent d'autres niveaux de complexité inhérents à l'utilisation de jeux de caractères et de systèmes multiples de transcription non standard (différents de l'Alphabet Phonétique International) ainsi qu'à la nécessité de tenir compte de la transcription de la parole dans une société primordialement orale, ce qui pose des problèmes d'encodage intéressant.

2. Le deuxième stade, inscrit quant à lui dans le champ des priorités directes de notre proposition, consiste à produire des référentiels normatifs de la TEI dans les langues locales et les faire valoir auprès des communautés académiques et professionnelles. Ces référentiels aideraient à traiter des cas de complexité spécifiques comme l'hybridation linguistique (dialectes locaux) des ressources numériques et la conservation d'une richesse culturelle orale et artistique millénaire. A ce titre, le problème des jeux de caractères n'est pas sans conséquence pour représenter les dialectes locaux, en grande partie parce que beaucoup de leurs aspects culturels n'ont pas été pris en compte dans l'élaboration de normes existantes (transcription des chiffres et des symboles, certaines formes de ligatures, diplomatique et alphabets anciens). Il existe, par exemple, de nombreuses propriétés de la langue arabe ou berbère, comme les propriétés tonales, la synonymie régionale et la voyellation ou vocalisation classique (écriture notariale) qui exigent des traitements spéciaux. Les normes en vigueur, en particulier la norme Unicode et encore moins les normes ISO 8859, ne tiennent pas compte de nombreux de ces aspects.

3. Le troisième stade - dans lequel nous sommes déjà engagés - est la création d'une communauté de pratique spécialisée dans le traitement des ressources spécifiques. Notons que la plupart de ces ressources sont potentiellement complexes et certaines de leurs caractéristiques nécessitent sans doute des modalités de balisage spécifiques. Cela signifie qu'un environnement dynamique est nécessaire pour la spécification de l'encodage de ces documents - un environnement dans lequel il est simple d'encoder des structures simples, mais dont les structures plus complexes peuvent aussi être encodées. Par conséquent, il est important d'avoir des spécifications qui peuvent être facilement étendues dès lors que de nouvelles fonctions intéressantes sont identifiés.
Ce qui nous intéresse dans la TEI c'est précisément sa dynamique collégiale ouverte sur la diversité linguistique non européenne (Japon, Chine, Corée), son éclectisme des disciplines de recherche (littérature, manuscrit, corpus oraux, recherche en art, linguistique…), mais aussi sa rigueur pour maintenir, enrichir et documenter des guidelines ouvertes sur la diversité mais à même d'assurer une interopérabilité de toutes les ressources produites.

## Résultats

Notre proposition de communication donnera un aperçu des travaux que nous entreprenons et des résultats que nous cherchons à communiquer.
Nous avons eu une première phase assez rapide d'appropriation de la TEI par des chercheurs maghrébins. Ils constitueront le relai vers une deuxième phase d'appropriation de la TEI par une communauté d'utilisateurs maghrébins plus large incluant dans les faits, des dizaines de thésards mais surtout des captations de savants des domaines non encore persuadés aujourd'hui de l'intérêt d'une TEI maghrébine. Il s'agit par exemple des spécialistes de la poétique arabe, des spécialistes de la linguistique berbère, des musicologues, des spécialistes du conte… Pour nous y prendre, la traduction en français et en arabe de la TEI P5, mais aussi le développement d'exemples et la mise en chantier d'une terminologie ou d'un glossaire TEI multilingue anglais/français/arabe nous paraît nécessaire.

Nous pensons aussi ouvrir des perspectives de recherche-action au sein de nouvelles communautés agissant sur le plan national et régional en synergie totale avec la dynamique internationale de la TEI. Nous l'avons déjà concrétisé par notre engagement dans un projet ayant pour objectif la production de ressources numériques franco-berbères, le projet de la Bibliothèque Numérique Franco-Berbère à financement francophone et d'un partenariat franco-maghrébin.

Bref, en nous inscrivant dans l'école de pensée des Humanités digitales et de la TEI, nous voulons explicitement donner non seulement une réalité tangible et numérique à nos travaux mais les rendre facilement échangeables, cumulables, améliorables, modifiables partout dans le monde. Nous voulons aussi sur un plan plus spécifique que nos travaux soient déjà facilement échangeables et cumulables entre nous et avec nos trois langues partenaires maghrébines (arabe, français, berbère) auxquelles il convient de rajouter l'anglais.

Il est évident que, hormis la question émergente du traitement et de la mise en patrimoine numérique, normalisé et interopérable, les spécialistes de ces patrimoines littéraires ont déjà largement exploré les méthodologies d'étude et de catalogage. De ce fait, cet article se limitera à discuter les seules questions posées par l'appropriation savante et professionnelle (bibliothèques et centres de recherche) des outils des Humanités digitales dans le contexte oriental. On s'intéressera entre autres aux problèmes culturels comparés opposant l'étude des manuscrits européens anciens par opposition au contexte arabe.